\documentclass[11pt,a4paper]{article}
\usepackage[dvips]{graphicx}
\title{
Vibrational energy relaxation (VER) of a CD stretching mode in cytochrome c
} 
\author{Hiroshi FUJISAKI\footnote{fujisaki@bu.edu}, Lintao BU\footnote{bult@bu.edu}, 
and John E. STRAUB\footnote{straub@bu.edu}
\\
\\
Department of Chemistry, Boston University, 590 Commonwealth Ave.,
\\
Boston, Massachusetts, 02215, USA}
\begin{document}

%\tableofcontents

%% To be entered at Wiley, for final production: title page information,
%% table of contents, preface and introduction, and Part titles, and index. 
%% See edbksamp.tex for examples of how to enter the commands.

% \footnote{text} will cause a footnote to appear at the bottom of the page.

%% 
%\title[Vibrational energy relaxation of a CD mode in cytochrome c]
%{Vibrational energy relaxation (VER) of a CD stretching mode in cytochrome c}
%\title[Controlling Quantum Chaos on Coarse-Grained Rabi Picture]
%{Controlling Quantum Chaos on Coarse-Grained Rabi Picture}
%\title[Coarse-Grained Rabi Oscillation and Control of Quantum Chaos]
%{Coarse-Grained Rabi Oscillation and Control of Quantum Chaos}

%% Please supply author names in upper and lower case within square
%% brackets and in uppercase in curly brackets, i.e.,
%% \author[The Author]{THE AUTHOR\footnote{Presently on leave at
%% NASA, Houston, Texas, USA.}}
%% You may use a footnote for additional information.

%\author[Hiroshi FUJISAKI, Lintao BU, and John E. STRAUB]
%{Hiroshi FUJISAKI, Lintao BU, and John E. STRAUB}
%\affil{Department of Chemistry, Boston University, 590 Commonwealth Ave.,
%Boston, Massachusetts, 02215, USA}
%\author[Lintao Bu]{Lintao BU}
%\affil{Department of Chemistry, Boston University, 590 Commonwealth Ave.,
%Boston, Massachusetts, 02215, USA}
%\author[John E. Straub]{John E. ST RAUB}
%\affil{Department of Chemistry, Boston University, 590 Commonwealth Ave.,
%Boston, Massachusetts, 02215, USA}

%\begin{document}
\maketitle

\begin{abstract}

We first review how to determine the rate  of
vibrational energy relaxation (VER) using 
perturbation theory.
We then apply those theoretical results to the problem 
of VER of a CD stretching mode in the protein cytochrome c.
We model cytochrome c in vacuum 
as a normal mode system with the lowest-order anharmonic coupling elements.
%which were calculated from the CHARMM potential. 
We find that, for the ``lifetime'' width parameter $\gamma=3 \sim 30$ cm$^{-1}$, 
the VER time is $0.2 \sim 0.3$ ps, which agrees rather well 
with the previous classical calculation using the quantum correction factor 
method, and is consistent with spectroscopic experiments by Romesberg's group.
%This result shows that the use of QCFs or the reduced model Hamiltonian 
%can be justified a posteriori to describe VER of the CD mode.
We decompose the VER rate into separate contributions from two modes, 
and find that the most significant contribution, 
which depends on the ``lifetime'' width parameter, 
comes from those modes most resonant with the CD vibrational mode.

\end{abstract}

%\tableofcontents 

\section{Introduction}

Vibrational energy relaxation (VER) is fundamentally important to 
our understanding of chemical reaction dynamics 
as it influences reaction rates significantly
\cite{BBS88}.
%which are often calculated (understood) by statistical theory like RRKM theory.
In general, estimating VER rates for selected modes in 
large molecules is a challenging problem because 
large molecules involve many degrees of freedom and, furthermore, 
quantum effects cannot be ignored \cite{Wolynes}.
If we assume a weak interaction between the ``system'' and the 
surrounding ``bath'',
however, we can derive an estimate of the VER rate through 
Fermi's golden rule \cite{Oxtoby81,BB94,KTF94,ES96}:
A VER rate is written as a Fourier transformation of a force-force correlation 
function.
Though it is not trivial to define and justify a separation of a system and a bath,
such a formulation has been successfully applied to many VER processes 
in liquids \cite{Hynes} 
and in proteins \cite{SS99}.

Here we apply such theories of VER to 
the problem of estimating the vibrational population relaxation 
time of a CD stretching mode, 
in short, a CD mode, in the protein cytochrome c \cite{BS03}.
(We will define the CD mode to be the system and the remainder 
of the protein to be the bath.)
Recently Romesberg's group succeeded in selectively deuterating a 
terimnal methyl group of a methionie residue in cytochrome c \cite{CJR01}.
The resulting CD mode has a frequency $\omega_S \simeq  2100$ cm$^{-1}$, 
which is located in a transparent region of the density of states of the protein.
As such, spectroscopic detection of this mode provides
clear evidence of the protein dynamics, including 
the VER of the CD vibrational mode.
Note that at room temperature ($T=300$ K) 
$\beta \hbar \omega_S \simeq 10$ where $\beta=1/(k_B T)$,
hence quantum effects are not negligible for this mode.

Let us mention a little more about cytochrome c (cyt c).
Cyt c is a protein known to exist in mitochondrial inner-membranes,
chloroplasts of plants, and bacteria \cite{cytc}.
Its functions are related to cell respiration \cite{Cell}, 
and cyt c, using its heme molecule, ``delivers'' an electron 
from cytochrome bc 1 to cytochrome oxidase -- two larger proteins 
both embedded in a membrane.
Recently it was also found that cyt c is released when apoptosis occurs \cite{apoptosis}. 
In this sense, cyt c governs the ``life and death'' of a cell.

The heme molecule in cyt c has a large oscillator strength, 
and serves as a good optical probe.
As a result, many spectroscopic experiments have been designed to clarify 
VER and the (un)folding properties of cyt c \cite{exp}.
Cyt c is often employed in numerical simulations \cite{simulation,BS03b}
because a high resolution structure was obtained \cite{structure}
and its simulation has become  feasible.
Attempts have also been made to characterize cyt c through 
ab initio (DFT) calculations \cite{DFT,ProteinDF}.

VER of the selected CD mode in the terminal methyl group of methionine (Met80)
was previously addressed by Bu and Straub \cite{BS03}:
They used equilibrium simulations for cyt c in water with the 
quantum-correction factor (QCF) method \cite{SP01}, and predicted that the 
VER time for the CD mode is on the order of 0.3 ps.
However, their results are approximate:
The use of the QCF method is not justified a priori,
and their analysis is based on a harmonic model for cyt c.
To extend that previous analysis, 
in this work, we model cyt c in vacuum 
as a normal mode system and include the lowest anharmonic coupling elements.
A similar analysis has been completed for another protein 
myoglobin by Kidera's group \cite{MMK03} 
and by Leitner's group \cite{Leitner01}.
Use of a reduced model Hamiltonian 
allows us to investigate the VER rate of the CD mode in cyt c more ``exactly''
and to move beyond the use of quantum correction factors and the 
harmonic approximation. 

This paper is organized as follows:
In Sec.~\ref{sec:formula},
we derive the principle VER formula employed in our work, 
and mention the related Maradudin-Fein formula. 
In Sec.~\ref{sec:application},
we apply those theoretical results for the rate of VER 
to the CD mode in cyt c,
and compare our results with the classical simulation by Bu and Straub,
and the experiments by Romesberg's group.
In Sec.~\ref{sec:summary}, we provide a summary of our results, and 
discuss further aspects of VER processes in proteins.

\section{Vibrational energy relaxation (VER)}
\label{sec:formula}

\subsection{Perturbation expansion for the interaction}

We begin with the von Neumann equation for the complete system
written as  
\begin{equation}
i \hbar \frac{d}{dt} \rho(t) =[{\cal H},\rho(t)].
\end{equation}
The interaction representation for the von Neumann equation is 
\begin{equation}
i \hbar \frac{d}{dt} \tilde{\rho}(t) =[\tilde{{\cal V}}(t),\tilde{\rho}(t)],
\end{equation}
where
\begin{equation}
{\cal H} ={\cal H}_0 + {\cal V} 
= {\cal H}_S +{\cal H}_B+ {\cal V} 
\end{equation}
and
\begin{equation}
\tilde{\rho}(t) \equiv 
e^{i {\cal H}_0 t/\hbar} \rho(t) e^{-i {\cal H}_0 t/\hbar},
\hspace{1cm}
\tilde{\cal V}(t) \equiv 
e^{i {\cal H}_0 t/\hbar} {\cal V} e^{-i {\cal H}_0 t/\hbar}.
\end{equation}
Here ${\cal H}_S$ is the system Hamiltonian representing 
a vibrational mode,
${\cal H}_B$ the bath Hamiltonian representing 
solvent or environmental degrees of freedom,
and ${\cal V}$ the interaction Hamiltonian describing the 
coupling between the system and the bath.
An operator with a tilde means the one in the interaction picture.
If we {\it assume} that ${\cal V}$ is small in some sense, 
we can carry out the perturbation expansion for ${\cal V}$ as
\begin{eqnarray}
\tilde{\rho}(t)
&=& \rho(0)+ \frac{1}{i \hbar} \int_0^t dt' 
[\tilde{\cal V}(t'), \tilde{\rho}(t')]
\nonumber
\\
&=&
\rho(0)+ \frac{1}{i \hbar} \int_0^t dt' 
[\tilde{\cal V}(t'), \rho(0)]
%\nonumber
%\\
%&&
+\frac{1}{(i \hbar)^2} \int_0^t dt' \int_0^{t'} dt''  
[\tilde{\cal V}(t'), [\tilde{\cal V}(t''),\rho(0)]]
+ \cdots.
\end{eqnarray}
Let us calculate the following probability: 
\begin{eqnarray}
P_{v}(t) 
&\equiv& {\rm Tr} \{ \rho_{v}  \rho(t) \}
= {\rm Tr} \{ \rho_{v}
e^{-i {\cal H}_0 t/\hbar} \tilde{\rho}(t) e^{i {\cal H}_0 t/\hbar}  \},
\\
\rho_{v} &\equiv& |v \rangle \langle v| \otimes 1_B,
\\
\rho(0) &=& 
\rho_S \otimes \rho_B
=
|v_0 \rangle \langle v_0| \otimes e^{-\beta {\cal H}_B}/Z_B,
\\
Z_B &=& {\rm Tr}_B \{ e^{-\beta {\cal H}_B} \},
\end{eqnarray}
where the initial state is assumed to be a direct product 
state of $\rho_S=|v_0 \rangle \langle v_0|$ and 
$\rho_B = e^{-\beta {\cal H}_B}/Z_B$.
Here $|v \rangle$ 
is the vibrational eigenstate for the system Hamiltonian ${\cal H}_S$,
i.e., ${\cal H}_S | v \rangle =E_v | v \rangle$.
The VER rate $\Gamma_{v_0 \rightarrow v}$ may be defined as follows:
\begin{equation}
\Gamma_{v_0 \rightarrow v} 
\equiv
\lim_{t \rightarrow \infty}
\frac{d}{dt} P_{v}(t). 
%\right |_{t \rightarrow \infty}.
\end{equation}
Note that the results derived from this definition are equivalent 
to those derived from Fermi's golden rule \cite{SR02}. 
Hence we refer to them as a Fermi's golden rule formula.

\subsection{General formula for VER}

First we notice that 
\begin{equation}
P_v(t)={\rm Tr} \{ \rho_{v}
e^{-i {\cal H}_0 t/\hbar} \tilde{\rho}(t) e^{i {\cal H}_0 t/\hbar}  \}
=
{\rm Tr} \{ \rho_{v} \tilde{\rho}(t)  \}
\end{equation}
as $\rho_v$ commutes with ${\cal H}_0$.
If we assume that $v \neq v_0$, then $\rho_v \rho(0)=0$.
Using this fact,
we obtain the lowest (second) order result  
\begin{eqnarray}
P_v(t) &\simeq&
\frac{1}{(i \hbar)^2} \int_0^t dt' \int_0^{t'} dt''  
{\rm Tr} \{ \rho_v 
[\tilde{\cal V}(t'), [\tilde{\cal V}(t''),\rho(0)]]
\}
\nonumber
\\
&=&
\frac{1}{\hbar^2} \int_0^t dt' \int_0^{t'} dt''  
{\rm Tr} 
\{ 
\rho_v \tilde{\cal V}(t') \rho(0) \tilde{\cal V}(t'')
+
\rho_v \tilde{\cal V}(t'') \rho(0) \tilde{\cal V}(t')
\}
\nonumber
\\
&=&
\frac{1}{\hbar^2} \int_0^t dt' \int_0^{t'} dt''  
[ e^{i \omega_{v_0 v}(t'-t'')} C(t'-t'')
%\nonumber
%\\
%&&
+
e^{i \omega_{v_0 v}(t''-t')} C(t''-t')]
\end{eqnarray}
where 
\begin{eqnarray}
C(t) &\equiv& 
\langle \tilde{\cal V}_{v_0 v} (t) {\cal V}_{v v_0}(0) \rangle
\equiv
{\rm Tr}_B \{ \rho_B
\tilde{\cal V}_{v_0 v}(t) {\cal V}_{v v_0}(0) \},
\\
\tilde{\cal V}_{v v_0}(t)
&=& \langle v | \tilde{\cal V}(t)| v_0 \rangle,
\\
\omega_{v_0 v} &=& (E_{v_0}-E_{v})/\hbar.
\end{eqnarray}
Hence the lowest order estimate of the VER rate is given by 
\begin{eqnarray}
\Gamma_{v_0 \rightarrow v}
&=& \lim_{t \rightarrow \infty}
\frac{1}{\hbar^2} \int_0^{t} dt''  
[ e^{i \omega_{v_0 v}(t-t'')} C(t-t'')
+
e^{i \omega_{v_0 v}(t''-t)} C(t''-t)
]
\nonumber
\\
&=&
%\frac{1}{\hbar^2} \int_0^{\infty} dt  
%[ e^{i \omega_{v_0 v}t} C(t)
%+
%e^{-i \omega_{v_0 v}t} C(-t)
%]
%=
\frac{1}{\hbar^2} \int_{-\infty}^{\infty} dt \, 
e^{i \omega_{v_0 v}t} C(t).
%=\frac{1}{\hbar^2} \tilde{C}(\omega_{v_0 v})
\end{eqnarray}
If we {\it assume} that a system variable $q$ is small in some sense,
the interaction Hamiltonian is expressed as   
\begin{equation}
{\cal V}= -q {\cal F}(\{ q_k \}, \{ p_k \})
\label{eq:assume}
\end{equation}
where $\{ q_k \}, \{ p_k \}$ are position and momentum variables 
for the bath. 
This ${\cal F}(\{ q_k \}, \{ p_k \})$ is %interpreted as 
a force applied to the system by the bath. 
Thus we finally obtain the following Fermi's golden rule formula \cite{Oxtoby81,BB94,KTF94,ES96}
\begin{eqnarray}
\Gamma_{v_0 \rightarrow v}
=\frac{|q_{v_0 v}|^2 }{\hbar^2} \int_{-\infty}^{\infty} dt \, 
e^{i \omega_{v_0 v}t} 
\langle \tilde{\cal F}(t) \tilde{\cal F}(0) \rangle
\label{Gamma}
\end{eqnarray}
where $q_{v_0 v}=\langle v_0|q|v \rangle$,
$\tilde{\cal F}(t)=e^{i {\cal H}_B t/\hbar} 
{\cal F} e^{-i {\cal H}_B t/\hbar}$, and  
$\langle \tilde{\cal F}(t) \tilde{\cal F}(0) \rangle= {\rm Tr}_B \{
\rho_B \tilde{\cal F}(t) \tilde{\cal F}(0) \}$.
In most situations, the transition from 
$v_0=1 $ to $v=0$ is considered. 
In such a case, $q_{10}=\sqrt{\hbar/2 m_S \omega_S}$,
%$\omega_{v_0 v}=\omega_S$,
%and the prefactor becomes $1/2 m_S \hbar \omega_S$ 
where $m_S$ is the system mass 
and $\omega_S=\omega_{10}$ is 
the system frequency in the harmonic approximation.
Hence 
\begin{eqnarray}
\Gamma_{1 \rightarrow 0}
=\frac{1}{2 m_S \hbar \omega_S} \int_{-\infty}^{\infty} dt \, 
e^{i \omega_{S}t} 
\langle \tilde{\cal F}(t) \tilde{\cal F}(0) \rangle.
\label{Gamma2}
\end{eqnarray}

\subsection{Use of a symmetrized auto-correlation function}

It is useful to define 
a {\it symmetrized} force-force correlation function as \cite{Oxtoby81,BB94,KTF94,ES96}
\begin{equation}
S(t)= \frac{1}{2} [\langle {\cal F}(t) {\cal F}(0) \rangle
+\langle {\cal F}(0) {\cal F}(t) \rangle].
\end{equation}
Since $S(t)$ is real and symmetric with respect to $t$, $S(t)=S(-t)$,
we consider it to be analogous to $S_{\rm cl}(t)$, 
the classical limit of the correlation function.
Hereafter we drop the tilde on ${\cal F}$ for simplicity.
By half-Fourier transforming $S(t)$ with the use of the 
relation $\langle {\cal F}(0) {\cal F}(t) \rangle
=\langle {\cal F}(t-i \beta \hbar) {\cal F}(0) \rangle$,
we have
\begin{eqnarray}
\int_0^{\infty} dt \, e^{i \omega t} S(t)
&=&
\frac{1}{2} \int_0^{\infty} dt \, e^{i \omega t} 
\langle {\cal F}(t) {\cal F}(0) \rangle
+
\frac{1}{2} \int_0^{\infty} dt \, e^{i \omega t} 
\langle {\cal F}(0) {\cal F}(t) \rangle
\nonumber
\\
&=&
\frac{1}{2} \int_0^{\infty} dt \, e^{i \omega t} 
\langle {\cal F}(t) {\cal F}(0) \rangle
+
\frac{1}{2} 
\int_0^{\infty} dt \, e^{i \omega t} 
\langle {\cal F}(t-i \beta \hbar) {\cal F}(0) \rangle
\nonumber
\\
&=&
\frac{1}{2} \int_0^{\infty} dt \, e^{i \omega t} 
\langle {\cal F}(t) {\cal F}(0) \rangle
+
\frac{1}{2} e^{-\beta \hbar \omega} 
\int_0^{\infty} dt \, e^{i \omega t} 
\langle {\cal F}(t) {\cal F}(0) \rangle
\nonumber
\\
&=&
\frac{1}{2} (1+e^{-\beta \hbar \omega})
\int_0^{\infty} dt \, e^{i \omega t} 
\langle {\cal F}(t) {\cal F}(0) \rangle
\end{eqnarray}
Taking the real parts of both sides, 
we have 
\begin{eqnarray}
\int_0^{\infty} dt \, \cos (\omega t) \, S(t)
= \frac{1}{4}(1+e^{-\beta \hbar \omega})
\int_{-\infty}^{\infty} dt \, e^{i \omega t} 
\langle {\cal F}(t) {\cal F}(0) \rangle
\end{eqnarray}
where we have used the fact that $S(-t)=S(t)$ is real and 
$\langle {\cal F}(t) {\cal F}(0) \rangle^*=
\langle {\cal F}(0) {\cal F}(t) \rangle  =
\langle {\cal F}(-t) {\cal F}(0) \rangle$.
Hence, Eq.~(\ref{Gamma2}) can be rewritten as \cite{ES96}
\begin{equation}
\Gamma_{1 \rightarrow 0}
=
\frac{1}{m_S \hbar \omega_S} 
\frac{2}{1+e^{-\beta \hbar \omega_S}}
\int_0^{\infty} dt \, \cos (\omega_S t)  \, S(t).
\end{equation}
%This form has a naive classical limit, i.e., 
%$S(t)$ can be regarded as a classical correlation 
%function calculated by classical mechanics.
Note that this expression diverges in the classical limit 
because $\Gamma_{1 \rightarrow 0} \propto 1/\hbar$.
According to Bader-Berne \cite{BB94}, 
to make contact with the classical limit,
we introduce another VER rate as  
\begin{eqnarray}
\frac{1}{T_1}
&=& (1-e^{-\beta \hbar \omega_S}) \Gamma_{1 \rightarrow 0}
\nonumber
\\
&=& 
2 C(\beta,\hbar \omega_S)
\int_0^{\infty} dt \, \cos (\omega_S t)  \, S(t)
\label{eq:VERrate}
\end{eqnarray}
where 
\begin{equation}
C(\beta,\hbar \omega_S)=
\frac{1}{m_S \hbar \omega_S} 
\frac{1-e^{-\beta \hbar \omega_S}}{1+e^{-\beta \hbar \omega_S}}.
\end{equation}
This is a final quantum expression,
which can be interpreted as an energy relaxation rate and 
be used to estimate the VER rate.\footnote{Although the experimental 
observable is $\Gamma_{1 \rightarrow 0}$, 
we note that $1/T_1 \simeq \Gamma_{1 \rightarrow 0}$ because  
$\beta \hbar \omega_S \gg 1$ for our case of a CD stretching mode.}

%\subsection{Classical limit and approximate methods
\subsection{Quantum correction factor method and other methods}

Though Eq.~(\ref{eq:VERrate}) is exact in a perturbative sense, 
it is demanding to calculate the {\it quantum mechanical} force autocorrelation 
function $S(t)$ even for small molecular systems. Hence, many 
computational schemes have been developed to approximate the 
quantum mechanical force autocorrelation function. 
% in terms of the 
%classical force autocorrelation function.

Skinner and coworkers advocated to use the quantum correction factor (QCF) method \cite{SP01},
which is the replacement of the above formula Eq.~(\ref{Gamma}) with
\begin{eqnarray}
\Gamma_{v_0 \rightarrow v}
= 
Q(\omega_S)
\frac{2 |q_{v_0 v}|^2}{\hbar^2} 
\int_0^{\infty} dt \, \cos (\omega_{v_0 v} t)  \, S_{\rm cl}(t)
%= \frac{Q(\omega_S)}{\beta \hbar \omega_S} \frac{1}{T_1^{\rm cl}}
\label{eq:qcf}
\end{eqnarray}
where $S_{\rm cl}(t)=\langle {\cal F}(t) {\cal F}(0) \rangle_{\rm cl}$ and 
the bracket means a classical ensemble average (not a quantum mechanical average).
This approach is very intuitive and easily applicable for large molecular systems
because one only needs to calculate the {\it classical} force autocorrelation 
function $S_{\rm cl}(t)$ multiplied by an appropriate QCF $Q(\omega_S)$. 
There exist several QCFs corresponding to different VER processes \cite{SP01}.
%, and these classic results have been unified and 
%substantially extended by Skinner and coworkers \cite{SP01}.

However, one challenge that arises in the application of the QCF method is 
that we do not know {\it a priori}
which VER process is dominant for the system considered.
Furthermore, it is possible that several VER processes compete \cite{SO98}.
Hence one must be careful in the application of the QCF method,
and Skinner and coworkers have provided a number of examples of how 
this can be accomplished.

In this paper,
we employ the harmonic approximation for the relaxing oscillator, 
and the vibrational relaxation time $T^{\rm QCF}_1$. 
Hence Eq.~(\ref{eq:qcf}) is transformed to
\begin{eqnarray}
\frac{1}{T_1^{\rm QCF}}
= 
\frac{Q(\omega_S)}{m_S \hbar \omega_S}
\int_0^{\infty} dt \, \cos (\omega_S t)  \, S_{\rm cl}(t)
= \frac{Q(\omega_S)}{\beta \hbar \omega_S} \frac{1}{T_1^{\rm cl}}
\label{eq:qcf2}
\end{eqnarray}
where we have introduced the classical VER rate $1/T_1^{\rm cl}$
\begin{eqnarray}
\frac{1}{T_1^{\rm cl}}
= \frac{\beta}{m_S} 
\int_0^{\infty} dt \, \cos (\omega_S t)  \, S_{\rm cl}(t)
\end{eqnarray}
which is the classical limit $\hbar \rightarrow 0$ of Eq.~(\ref{eq:VERrate}).
%where $S_{\rm cl}(t)$ is a classical limit of $S(t)$.
This result can also be derived from a classical theory of Brownian motion,
and is known as the Landau-Teller-Zwanzig (LTZ) formula.

Alternatively, one may calculates $S(t)$ {\it itself} systematically using
controlled approximations.
Calculating a correlation function for large systems has a long 
history in chemical physics \cite{Miller01}, including recent applications
 to VER processes in liquid \cite{SG03a,Rossky}. 
The vibrational self-consistent field (VSCF) method \cite{RGER95} 
will also be useful in this respect.

%Since the full quantum calculation of $S(t)$ is demanding 
%even for the smallest molecules, the LTZ formula is often invoked to estimate 
%VER rates in the literature \cite{SS99}. 
%However, it is not certain that 
%such a classical calculation can accurately describe actual
%VER processes, particularly in the region where $\beta \hbar \omega_S \gg 1$. 
%When a vibrational mode 
%with a high frequency is involved, one anticipates that 
%quantum effects will contribute to the process of VER.

On the other hand, if we {\it approximate} ${\cal F}$ as a simple 
function of $\{ q_k \}, \{ p_k \}$, we can calculate $S(t)$ 
rather easily and, in a sense, more ``exactly.'' 
In the next section, we explore such an approach.

\subsection{Approximations for the force-force correlation function}

\subsubsection{Taylor expansion of the force}
\label{sec:taylor}

We can formally Taylor-expand the force
as a function of the bath variables $\{ q_k \}, \{ p_k \}$:
\begin{equation}
{\cal F}(\{ q_k \}, \{ p_k \})= 
\sum_k A^{(1)}_k q_k 
+
\sum_k B^{(1)}_k p_k 
+
\sum_{k,k'} A^{(2)}_{k,k'} q_k q_{k'} 
+
\sum_{k,k'} B^{(2)}_{k,k'} p_k p_{k'}
+ \cdots 
\label{eq:taylor}
\end{equation}
where the expansion is often truncated in the literature following 
the first term.
Depending on the system-bath interaction considered, 
higher order coupling including the third and fourth terms can be relevant.
For example, the fourth term appears in benzene to 
represent the interaction between the CH stretch
and CCH wagging motion \cite{SHR84} through the Wilson G matrix \cite{Wilson}.
In the case of a CD stretching mode in cyt c, as discussed below, 
or a CN$^-$ stretching mode in water \cite{SO98}, 
the third term is relevant for VER.

\subsubsection{Contribution from the first term}

If the first term 
$\sum_k A^{(1)}_k q_k$ 
is dominant in the force, 
then the force-force correlation function becomes
\begin{equation}
\langle {\cal F}(t) {\cal F}(0) \rangle
= \sum_{k,k'} A_k^{(1)} A_{k'}^{(1)} 
\langle q_k(t) q_{k'}(0) \rangle
=\sum_k \frac{\hbar (A_k^{(1)})^2}{2 \omega_k} 
[(n_k+1) e^{-i \omega_k t}+ n_k e^{i \omega_k t}]
\end{equation}
where we have used 
\begin{equation}
q_k(t)=\sqrt{\frac{\hbar}{2 \omega_k}} 
(a_k e^{-i \omega_k t}+a^{\dagger}_k e^{i \omega_k t})
\end{equation}
and $\langle a^{\dagger}_k a_{k'} \rangle = n_k \delta_{k,k'}$ with 
$n_k=1/(e^{\beta \hbar \omega_k}-1)$
because 
\begin{equation}
\rho_B \propto e^{-\beta {\cal H}_B} = e^{-\beta 
\sum_k \hbar \omega_k (a_k^{\dagger} a_k +1/2)}.
\end{equation}
Here we have {\it assumed} that the bath Hamiltonian is an 
ensemble of harmonic oscillators:
\begin{equation}
{\cal H}_B = 
\sum_k \hbar \omega_k (a_k^{\dagger} a_k +1/2)
=
\sum_k \left( 
\frac{p_k^2}{2}+\frac{\omega_k^2}{2} q_k^2 
\right)
\end{equation}
where $\omega_k$ is the $k$-th mode frequency for the bath, 
and 
\begin{equation}
p_k =-i \sqrt{\frac{\hbar \omega_k}{2}} (a_k -a_k^{\dagger}).
\end{equation}
Thus we obtain 
\begin{equation}
S(t)=\sum_k \frac{\hbar (A_k^{(1)})^2}{2 \omega_k} 
(2 n_k+1) \cos \omega_k t
\end{equation}
and 
\begin{eqnarray}
\frac{1}{T_1}
= 
\pi \hbar C(\beta,\hbar \omega_S)
\sum_k \frac{(A_k^{(1)})^2}{\omega_k} 
(2 n_k+1) [\delta(\omega_S -\omega_k)+ \delta(\omega_S+ \omega_k)].
%\nonumber
%\\
%&\simeq& 
%\pi \hbar C(\beta,\hbar \omega_S)
%\rho(\omega_S) D^{(1)}(\omega_S) [2 n(\omega_S) +1]
\label{eq:cont}
\end{eqnarray}
%where we have assumed that $\omega_S,\omega_k >0$ and defined 
%$D^{(1)}(\omega)=[A^{(1)}(\omega)]^2/\omega$ as a continuum 
%limit of $(A^{(1)}_k)^2 /\omega_k$.
The contribution from the second term 
$\sum_k B^{(1)}_k p_k$ 
is calculated in the same way.
%In the {\bf high frequency (low temperature)} regimes, i.e.,
%when $n_k \ll 1$,
%this turns out to be 
%\begin{equation}
%\Gamma_{v}^{(2)}
%=\frac{\pi |q_{v_0 v}|^2 }{\hbar} 
%\sum_k \frac{(A_k^{(1)})^2}{m_k \omega_k} 
%\delta(\omega_S -\omega_k)
%= \frac{\pi |q_{v_0 v}|^2 }{\hbar} J(\omega_S)
%\end{equation}
%where $J(\omega_S)$ is the {\bf spectral density} for the bath.

%\subsubsection{Contribution from the second term}

%If the second term 
%$\sum_k B^{(1)}_k p_k$ 
%is dominant, then 
%\begin{equation}
%\langle {\cal F}(t) {\cal F}(0) \rangle
%= \sum_{k,k'} B_k^{(1)} B_{k'}^{(1)} 
%\langle p_k(t) p_{k'}(0) \rangle
%=\sum_k \frac{\hbar (B_k^{(1)})^2 m_k \omega_k}{2} 
%[(n_k+1) e^{-i \omega_k t}+ n_k e^{i \omega_k t}]
%\end{equation}
%where we have used 
%\begin{equation}
%p_k(t)=-i \sqrt{\frac{\hbar m_k \omega_k}{2}} 
%(a_k e^{-i \omega_k t}-a^{\dagger}_k e^{i \omega_k t})
%\end{equation}
%and the other stuffs as the same as the above.
%Thus we obtain 
%\begin{equation}
%\Gamma_{v}^{(2)}
%=\frac{\pi |q_{v_0 v}|^2 }{\hbar} 
%\sum_k (B_k^{(1)})^2 m_k \omega_k
%[(n_k+1) \delta(\omega_S -\omega_k)+ n_k \delta(\omega_S+ \omega_k)] 
%\end{equation}

\subsubsection{Contribution from the third term}

If the third term $\sum_{k,k'} A^{(2)}_{k,k'} q_k q_{k'}$  
is dominant in the force, then 
\begin{eqnarray}
\langle {\cal F}(t) {\cal F}(0) \rangle
&=& \sum_{k,k',k'',k'''} A_{k,k'}^{(2)} A_{k'',k'''}^{(2)} 
\langle q_k(t) q_{k'}(t) q_{k''}(0) q_{k'''}(0) \rangle
\nonumber
\\
&=&
R_{--}(t)+R_{++}(t)+R_{+-}(t)
\end{eqnarray}
where 
\begin{eqnarray}
R_{--}(t)
&=&
\frac{\hbar^2}{4}
\sum_{k,k',k'',k'''} 
D^{(2)}_{k,k',k'',k'''}
\langle a_k a_{k'} a^{\dagger}_{k''} a^{\dagger}_{k'''} \rangle
e^{-i(\omega_k +\omega_{k'})t}, 
\\
R_{++}(t)
&=&
\frac{\hbar^2}{4}
\sum_{k,k',k'',k'''} 
D^{(2)}_{k,k',k'',k'''}
\langle a^{\dagger}_{k} a^{\dagger}_{k'} a_{k''} a_{k'''} \rangle
e^{i(\omega_k +\omega_{k'})t}, 
\\
R_{+-}(t)
&=&
\frac{\hbar^2}{4}
\sum_{k,k',k'',k'''} 
D^{(2)}_{k,k',k'',k'''}
\left[
\langle 
a_{k} a^{\dagger}_{k'}
(a^{\dagger}_{k''} a_{k'''}+ a_{k''} a^{\dagger}_{k'''})
\rangle
e^{-i(\omega_k -\omega_{k'})t}
\right.
\nonumber
\\
&&
+
\left.
\langle 
a^{\dagger}_{k} a_{k'}
(a^{\dagger}_{k''} a_{k'''}+ a_{k''} a^{\dagger}_{k'''})
\rangle
e^{i(\omega_k -\omega_{k'})t}
\right] 
\end{eqnarray}
with 
\begin{equation}
D^{(2)}_{k,k',k'',k'''}
=
\frac{A_{k,k'}^{(2)} A_{k'',k'''}^{(2)}} 
{\sqrt{\omega_k \omega_{k'} \omega_{k''} \omega_{k'''}}}. 
\end{equation}
Using the following 
\begin{eqnarray}
\langle a_k a_{k'} a^{\dagger}_{k''} a^{\dagger}_{k'''} \rangle
&=&
(1+n_k)(1+n_{k'})(\delta_{kk''}\delta_{k'k'''}+\delta_{kk'''}\delta_{k'k''}),
\\
\langle a^{\dagger}_{k} a^{\dagger}_{k'} a_{k''} a_{k'''} \rangle
&=&
n_k n_{k'}(\delta_{kk''}\delta_{k'k'''}+\delta_{kk'''}\delta_{k'k''}),
\\
\langle 
a_{k} a^{\dagger}_{k'}
(a^{\dagger}_{k''} a_{k'''}+ a_{k''} a^{\dagger}_{k'''})
\rangle
&=&
(1+n_k)(1+2 n_{k''}) \delta_{kk'}\delta_{k''k'''}
\nonumber
\\
&&
+
(1+n_k) n_{k'}(\delta_{kk''}\delta_{k'k'''}+\delta_{kk'''}\delta_{k'k''}),
\\
\langle 
a^{\dagger}_{k} a_{k'}
(a^{\dagger}_{k''} a_{k'''}+ a_{k''} a^{\dagger}_{k'''})
\rangle
&=&
n_k (1+2 n_{k''}) \delta_{kk'}\delta_{k''k'''}
\nonumber
\\
&&
+
n_k(1+n_{k'})(\delta_{kk''}\delta_{k'k'''}+\delta_{kk'''}\delta_{k'k''}),
\end{eqnarray}
we have 
\begin{eqnarray}
R_{--}(t)
&=&
\frac{\hbar^2}{2}
\sum_{k,k'} 
D^{(2)}_{k,k',k,k'}
(1+n_k)(1+n_{k'})
e^{-i(\omega_k +\omega_{k'})t}, 
\\
R_{++}(t)
&=&
\frac{\hbar^2}{2}
\sum_{k,k'} 
D^{(2)}_{k,k',k,k'}
n_k n_{k'}
e^{i(\omega_k +\omega_{k'})t}, 
\\
R_{+-}(t)
&=&
\langle {\cal F}(0) \rangle^2
%\frac{\hbar^2}{4}
%\sum_{k,k'} 
%D^{(2)}_{k,k,k',k'}
%(1+2 n_k)(1+2 n_{k'})
%\nonumber
%\\
%&+&
+
\hbar^2
\sum_{k,k'} 
D^{(2)}_{k,k',k,k'}
(1+n_k)n_{k'}
e^{-i(\omega_k -\omega_{k'})t}
\end{eqnarray}
where we have used $A^{(2)}_{k,k'}=A^{(2)}_{k',k}$, and 
\begin{equation}
\langle {\cal F}(0) \rangle^2
=
\frac{\hbar^2}{4} \sum_{k,k'} 
D^{(2)}_{k,k,k',k'} (1+ 2 n_k)(1+ 2 n_{k'}).
\end{equation}
%The average of the force itself is 
%\begin{equation}
%\langle {\cal F}(t) \rangle
%=
%\frac{\hbar}{2}
%\sum_k 
%\frac{A^{(2)}_{k,k}}{m_k \omega_k} 
%(1+2 n_k) 
%=\langle {\cal F}(0) \rangle.
%\end{equation}
%Note that, in this case, 
%$\langle {\cal F}(t) {\cal F}(0) \rangle
%\neq \langle \delta{\cal F}(t) \delta{\cal F}(0) \rangle$.
%But, to compare with classical results, we should take 
%$\langle \delta{\cal F}(t) \delta{\cal F}(0) \rangle$.
%Actually $\langle {\cal F}(t) \rangle \langle {\cal F}(0) \rangle$ 
%= constant only gives a VER rate at $\omega_S=0$, thus it does not 
%make any difference for almost all cases.
Hence we obtain
\begin{eqnarray}
S(t)
=
\sum_{k,k'} 
\left[
\zeta_{k,k'}^{(+)} 
\cos(\omega_k+\omega_{k'})t 
+\zeta_{k,k'}^{(-)} 
\cos(\omega_k-\omega_{k'})t 
\right]
+\langle {\cal F}(0) \rangle^2
\label{eq:ffc2}
\end{eqnarray}
and 
\begin{eqnarray}
\frac{1}{T_1}
=
\pi C(\beta,\hbar \omega_S)
\sum_{k,k'} 
\left \{ \zeta_{k,k'}^{(+)} 
[\delta(\omega_k+\omega_{k'}-\omega_S)
+\delta(\omega_k+\omega_{k'}+\omega_S)]
\right.
\nonumber
\\
\left.
+
\zeta_{k,k'}^{(-)} 
[\delta(\omega_k-\omega_{k'}-\omega_S)
+\delta(\omega_k-\omega_{k'}+\omega_S)]
\right \}
\label{eq:VERrate2}
\end{eqnarray}
where we have assumed $\omega_S \neq 0$ and defined 
\begin{eqnarray}
\zeta_{k,k'}^{(+)} 
&=& \frac{\hbar^2}{2} 
D^{(2)}_{k,k',k,k'} (1+ n_k + n_{k'} + 2 n_k n_{k'}),
\\
\zeta_{k,k'}^{(-)} 
&=& \frac{\hbar^2}{2} 
D^{(2)}_{k,k',k,k'} (n_k + n_{k'} + 2 n_k n_{k'}).
\end{eqnarray}
Though its appearance is rather different, 
the formula Eq.~(\ref{eq:VERrate2}) 
is equivalent to that derived by Kenkre, Tokmakoff, and Fayer \cite{KTF94}
as well as by Shiga and Okazaki \cite{SO98}. 
%with a different method (path-integral). 
There is also a similar result known as the Maradudin-Fein formula \cite{MF62}
\begin{eqnarray}
W &=& W_{\rm decay} + W_{\rm coll},
\label{eq:MFformula}
\\
W_{\rm decay} 
&=&
\frac{\pi \hbar}{2 m_S \omega_S} 
\sum_{k,k'} \frac{(A_{k,k'}^{(2)})^2}{\omega_k \omega_{k'}}
(1+n_k +n_{k'}) \delta(\omega_S -\omega_k-\omega_{k'}),
\\
W_{\rm coll}
&=&
\frac{\pi \hbar}{m_S \omega_S} 
\sum_{k,k'} \frac{(A_{k,k'}^{(2)})^2}{\omega_k \omega_{k'}}
(n_k -n_{k'}) \delta(\omega_S +\omega_k-\omega_{k'}),
\end{eqnarray}
which has been utilized to describe VER processes in glasses \cite{FA96}
and in proteins by Leitner's group \cite{Leitner01}.
%Actually %In the quantum regime, $\beta \hbar \omega_S \gg 1$, 
%the two formulas, Eqs.~(\ref{eq:VERrate2}) and (\ref{eq:MFformula}), 
%behave very similarly as shown below for the case of the CD mode in cyt c.
As was demonstrated in \cite{KTF94}, this formula is also equivalent 
to Eq.~(\ref{eq:VERrate2}); in the following we make use of  
Eq.~(\ref{eq:VERrate2}). 
A problem with this formula is that we cannot take its 
continuum limit in the case of finite systems like proteins.
%as easily as Eq.~(\ref{eq:cont}).
As a remedy, a {\it width parameter} related to the vibrational lifetime 
is usually introduced leading to a definite value for the VER rate.
We will discuss this problem in Sec.~\ref{sec:width}.

\section{Application to a CD stretching mode in cytochrome c}
\label{sec:application}

\subsection{Definition of system and bath}

We take horse heart cytochrome c (cyt c) as an example of how one may
estimate the rate of VER for selected modes in proteins.
We use the CHARMM program \cite{CHARMM} to describe the force field,
to minimize the structure, and to calculate the 
normal modes for the system.
Starting from the 1HRC structure for cyt c 
in Protein Data Bank (PDB) \cite{PDB} one hydrogen atom of the terminal methyl 
group of Met80 was deuterated.  
The energy of the protein structure was minimized in vacuum using 
the conjugate gradient algorithm.
We diagonalized the 
hessian matrix (second derivatives of the potential) 
for that mechanically stable configuration of the protein:
\begin{equation}
K_{ij}= 
\frac{\partial^2 V_{\rm CHARMM}}{\partial \bar{x}_i \partial \bar{x}_j}
=
\frac{1}{\sqrt{m_i m_j}}
\frac{\partial^2 V_{\rm CHARMM}}{\partial x_i \partial x_j}
\end{equation}
where $V_{\rm CHARMM}$ is the CHARMM potential, and
$\bar{x}_i=\sqrt{m_i} x_i$ are mass weighted Cartesian coordinates.
The number of atoms in cyt c is 1745 (myoglobin has 2475 atoms), 
so the hessian matrix is 5235 $\times$ 5235,
and its diagonalization was readily accomplished
using the {\tt vibran} facility in CHARMM \cite{CHARMM}.

\begin{figure}[htbp]
\hfill
\begin{center}
%\begin{minipage}{.42\linewidth}
\includegraphics[scale=1.5]{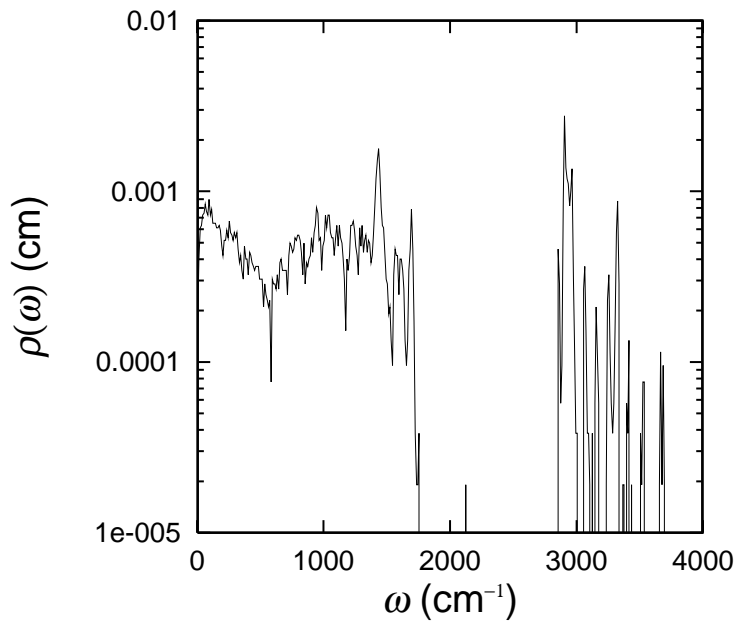}
\end{center}
\caption{
%{\sf 
Density of states for cytochrome c in vacuum.
%}
}
\label{fig:dos}
\end{figure}

\begin{figure}[htbp]
\hfill
\begin{center}
\begin{minipage}{.32\linewidth}
\includegraphics[scale=0.8]{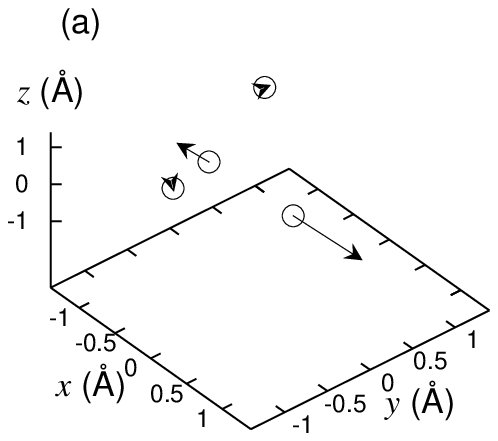}
\end{minipage}
\begin{minipage}{.32\linewidth}
\includegraphics[scale=0.8]{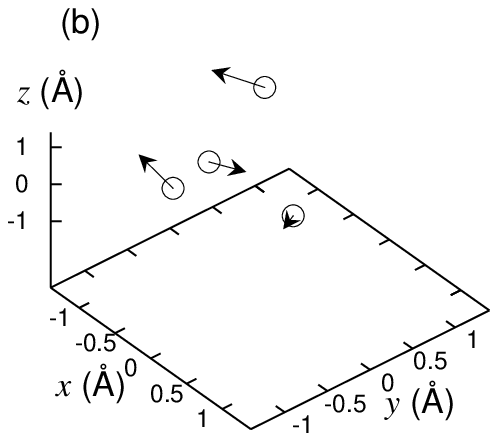}
\end{minipage}
\begin{minipage}{.32\linewidth}
\includegraphics[scale=0.8]{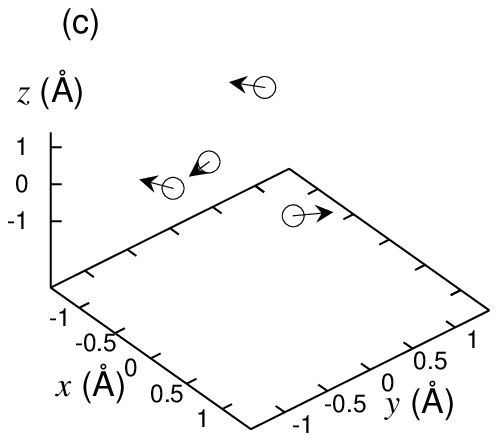}
\end{minipage}
\end{center}
\caption{
%{\sf 
Normal modes of cytochrome c in vacuum.
(a) 4357th mode (CD mode) with $\omega=2129.1$ cm$^{-1}$,
(b) 3330th mode with $\omega=1330.9$ cm$^{-1}$,
(c) 1996th mode with $\omega=829.9$ cm$^{-1}$.
Only vectors on the terminal methyl group of Met80 in cyt c 
are depicted.
%}
}
\label{fig:modes}
\end{figure}

The result of this calculation was the density of states (DOS) for the system as shown in 
Fig.~\ref{fig:dos}. The DOS consists of three regions: 
(1) below around 1700 cm$^{-1}$, 
(2) from around 1700 cm$^{-1}$ to around 2800 cm$^{-1}$,   
(3) above around 2800 cm$^{-1}$.
The first region corresponds to rotational and torsional motions 
of the protein, and the third to bond stretching motions such as CH bonds.
The second is rather ``transparent'' but one can observe one mode 
localized around the CD bond stretching mode in Met80 with 
frequency 2129.1 cm$^{-1}$ as shown in Fig.~\ref{fig:modes} (a).
Hence we refer to this as a CD stretching mode, or CD mode; 
the dynamics of which is the focus of our study.

At this level of description, the system is an ensemble of 
harmonic oscillators, i.e., normal modes.
Since we are interested in VER of the CD mode,
we represent it as a system 
\begin{equation}
{\cal H}_S
=\frac{p_{CD}^2}{2}+\frac{\omega_{CD}^2}{2}q_{CD}^2
\end{equation}
while other degrees of freedom are treated as a bath
\begin{equation}
{\cal H}_B
=
\sum_k 
\left(
\frac{p_k^2}{2}+\frac{\omega_k^2}{2}q_k^2
\right).
\end{equation}
The interaction between the system and bath is described by
the interaction Hamiltonian
\begin{equation}
{\cal V}= {\cal H}_{\rm cytc} -{\cal H}_S-{\cal H}_B
\end{equation}
where ${\cal H}_{\rm cytc}$ is the Hamiltonian for the full cyt c protein.
We will discuss the content of ${\cal V}$ in the following section.
%+q_{CD} \sum_{m,n} C_{mn} q_m q_n.
%\label{eq:ham}
%\end{equation}

\subsection{Calculation of the coupling constants}

As in Eq.~(\ref{eq:assume}), we {\it assume} that the interaction Hamiltonian 
is of the form
\begin{equation}
{\cal V}=-q_{CD} {\cal F}
\end{equation}
and Taylor expand the force as Eq.~(\ref{eq:taylor}).
The first and second terms do not appear because this is 
a normal mode expansion, and the fourth term does not appear as 
the original coordinates are Cartesian coordinates.
As in the first approximation,
we take the force to be 
\begin{equation}
{\cal F} = \sum_{k,k'} A^{(2)}_{k,k'} q_k q_{k'}.
\end{equation}
The coupling coefficients $A^{(2)}_{k,l}$ are calculated as 
\begin{equation}
A^{(2)}_{k,l} = 
-\frac{1}{2} \frac{\partial^3 V}{\partial q_{CD} \partial q_k \partial q_l}.
\end{equation}
A problem arises: How does one calculate these coupling coefficients?
The most direct approach is to use a finite difference method: 
\begin{equation}
A^{(2)}_{k,l} 
%&=& \frac{1}{2} \frac{\partial^3 V}{\partial q_{CD} \partial q_m \partial q_n}
%\nonumber
%\\
\simeq
-\frac{1}{2} \frac{
V_{+++}-V_{-++}-V_{+-+}-V_{++-}+V_{--+}+V_{-+-}+V_{+--}-V_{---}}
{(2 \Delta q_{CD})(2 \Delta q_k)(2 \Delta q_l)}
\label{eq:finite}
\end{equation}
where $V_{\pm \pm \pm} = V(\pm \Delta q_{CD}, \pm \Delta q_k, \pm \Delta q_l)$.
%I took $\Delta q_{CD}=0.02$ (AKMA) for the calculation.
However, this is rather cumbersome. 
Instead, we use the approximation \cite{Leitner01,SO98}:
\begin{equation}
A^{(2)}_{k,l} 
\simeq 
-\frac{1}{2} \sum_{ij} U_{ik}U_{jl}
\frac{K_{ij}(\Delta q_{CD})-K_{ij}(-\Delta q_{CD})}{2\Delta q_{CD}}
\label{eq:formula}
\end{equation}
where $U_{ik}$ is an orthogonal matrix that diagonalizes 
the hessian matrix at the mechanically stable structure $K_{ij}$,
and $K_{ij}(\pm \Delta q_{CD})$ is a hessian matrix 
calculated at a shifted structure along the direction of the CD mode
with a shift $\pm \Delta q_{CD}$.
This expression is approximate but readily implemented 
using the CHARMM facility to compute the hessian matrix.
A comparison between Eqs.~(\ref{eq:finite}) and (\ref{eq:formula}) 
is made in Table \ref{tab:comparison}.
We also examined the convergence of the results by
changing $\Delta q_{CD}$, and found that $\Delta q_{CD}=0.02 {\rm \AA}$
is sufficient for the following calculations.

\begin{table}[htbp]
\caption{
%{\sf 
Comparison between the finite difference method 
Eq.~(\ref{eq:finite}) and the formula Eq.~(\ref{eq:formula}).
We have used $\Delta q_{CD}=0.02 $ \AA,
and $A^{(2)}_{k,l}$ is given in kcal/mol/\AA$^3$.
Note that $(k,l)$ are mode numbers, not wavenumbers.
%}
}
\hfill
\begin{center}
\begin{tabular}{c|c|c}
\hline \hline
$(k,l)$ & formula, Eq.~(\ref{eq:formula})& Finite difference \\ \hline \hline
(3330, 1996) & 22.3 & 22.4 \\ \hline
(3330, 4399) & -29.6 & -29.5 \\ \hline
(3327, 1996) & -5.7 & -5.8 \\ \hline
(1996, 678) & 0.64 & 0.63 \\ 
\hline
\end{tabular}
\end{center}
\label{tab:comparison}
\end{table}

\begin{figure}[htbp]
\hfill
\begin{center}
\begin{minipage}{.42\linewidth}
\includegraphics[scale=1.0]{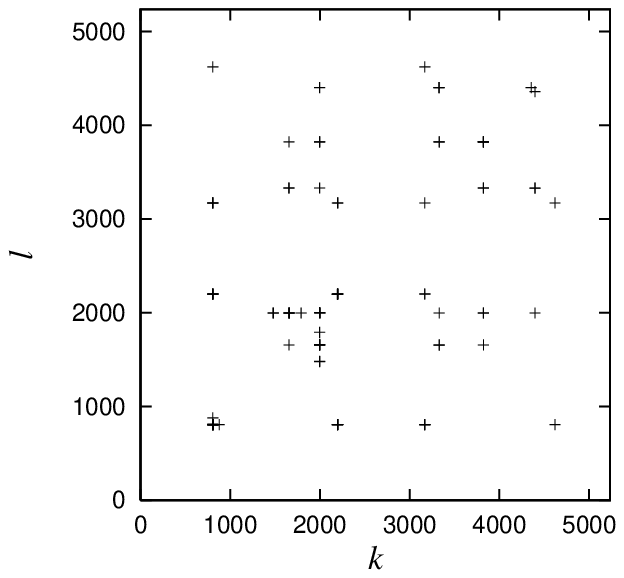}
\end{minipage}
\hspace{1cm}
\begin{minipage}{.42\linewidth}
\includegraphics[scale=1.0]{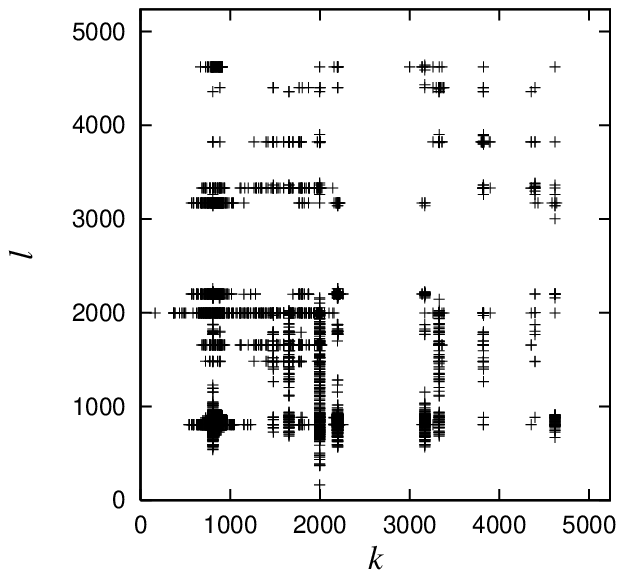}
\end{minipage}
\end{center}
\caption{
%{\sf 
Distribution of the coupling elements.
Left: $50.0> |A^{(2)}_{k,l}| > 5.0$, 
Right: $5.0 > |A^{(2)}_{k,l}| > 0.5$. 
The value of $A^{(2)}_{k,l}$ 
are given in units of kcal/mol/\AA$^3$.
Note that $(k,l)$ are mode numbers, not wavenumbers.
%}
}
\label{fig:couple}
\end{figure}

\begin{figure}[htbp]
\hfill
\begin{center}
\includegraphics[scale=1.5]{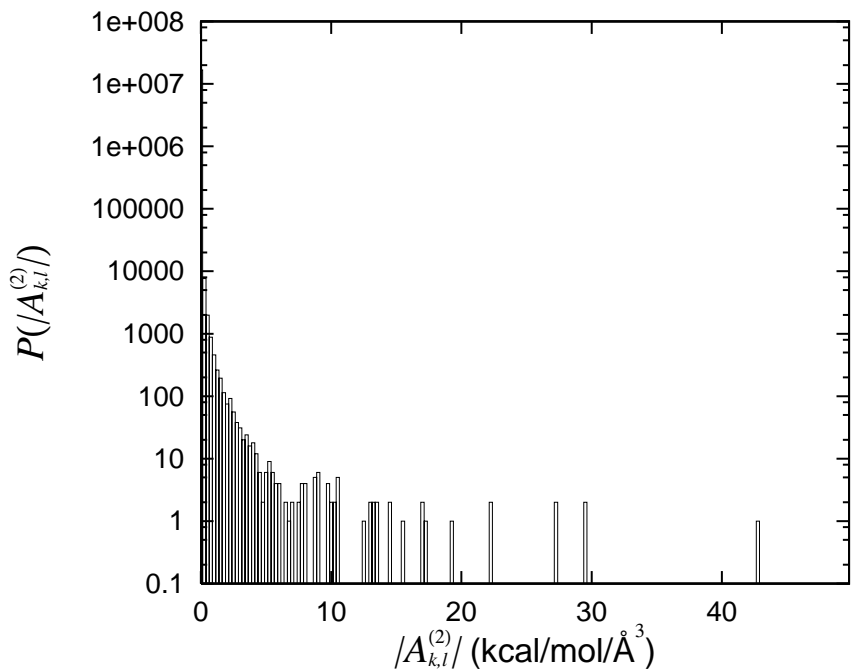}
\end{center}
\caption{
%{\sf 
Histogram for the amplitude of the coupling elements.
%The unit for $A^{(2)}_k,l}$ 
%is kcal/mol/\AA$^3$.
%}
}
\label{fig:dist}
\end{figure}

The numerical results for the coupling elements 
are shown in Fig.~\ref{fig:couple}. 
The histogram for the elements is  
shown in Fig.~\ref{fig:dist}. 
As one can see from these figures, 
most of the elements are small, 
while the largest coupling elements are rather large. 
See Table \ref{tab:largest}.
Note that the combination (3330, 1996) is 
particularly significant for the CD mode because 
it approximately satisfies the {\it resonant condition} \cite{MMK03}:
\begin{equation}
|\omega_{CD}-\omega_k -\omega_l| \ll {\cal O}(|A^{(2)}_{k,l}|).
\end{equation}
As shown in Fig.~\ref{fig:modes} (b), (c), 
these modes are localized near the terminal methyl group 
of Met80 as well as the CD mode.
In such a case, resonant energy transfer (Fermi resonance) is expected 
as shown by Moritsugu-Miyashita-Kidera \cite{MMK03}.
We have observed similar behavior in cyt c 
when the CD mode was excited and the 
energy immigration to other normal modes facilitated 
by resonance was followed.

\begin{table}[htbp]
\caption{
%{\sf 
The largest coupling elements. 
The value of $A^{(2)}_{k,l}$ 
are given in kcal/mol/\AA$^3$.
Note that $(k,l)$ are mode numbers, not wavenumbers.
%}
}
\hfill
\begin{center}
\begin{tabular}{c|c}
\hline \hline
$(k,l)$ & $|A^{(2)}_{k,l}|$ \\ \hline \hline
(1996, 1996) & 42.9  \\ \hline
(4399, 3330) & 29.6 \\ \hline
(4622, 3170) & 27.3  \\ \hline
(3330, 1996) & 22.3 \\ 
\hline
\end{tabular}
\end{center}
\label{tab:largest}
\end{table}

\subsection{Assignment of the ``lifetime'' parameter}
\label{sec:width}

We cannot directly evaluate 
Eq.~(\ref{eq:VERrate2}) 
as it contains delta functions.
Evaluation of this expression for 
a finite system like a protein leads to a null result.
To circumvent this problem,
we ``thaw'' the delta function $\delta(x)$ as 
\begin{equation}
\delta(x)=\frac{1}{\pi} \frac{\gamma}{\gamma^2 +x^2}
\end{equation}
using a width parameter $\gamma$.
Physically this means that each normal mode should 
have a lifetime $\simeq 1/\gamma$ due to 
coupling to other degrees of freedom, i.e., the 
surrounding environment including water
%\bibitem{KLC04}
(or we might be able to interpret $1/\gamma$ as a time resolution).
%See, e.g.,  K.~Kwac, H.~Lee, and M.~Cho,
%``Non-Gaussian statistics of amide I mode frequency fluctuation
%of $N$-methylacetamide in methanol solution:
%Linear and nonlinear vibrational spectra,''
%J.~Chem.~Phys.~{\bf 120}, 1477 (2004).
%\cite{KLC04}.
%}
It is difficult to derive $\gamma$ from first principles,
so we take it to be 
a phenomenological parameter as in the literature \cite{Leitner01,FA96}.

As a result, the VER rate, Eq.~(\ref{eq:VERrate2}), for the CD mode becomes
\begin{eqnarray}
\frac{1}{T_1}
= 
C(\beta,\hbar \omega_S)
\sum_{k,k'} 
\left[
\frac{\gamma \zeta^{(+)}_{k,k'}}{\gamma^2+(\omega_k+\omega_{k'}-\omega_S)^2}
+
\frac{\gamma \zeta^{(+)}_{k,k'}}{\gamma^2+(\omega_k+\omega_{k'}+\omega_S)^2}
\nonumber
\right.
\\
\left.
+
\frac{\gamma \zeta^{(-)}_{k,k'}}{\gamma^2+(\omega_k-\omega_{k'}-\omega_S)^2}
+
\frac{\gamma \zeta^{(-)}_{k,k'}}{\gamma^2+(\omega_k-\omega_{k'}+\omega_S)^2}
\right].
\label{eq:VERrate3}
\end{eqnarray}
%where we have assumed $e^{-\beta \hbar \omega_S} \simeq 0$.
%This is the case at $T \simeq 300$K because $\beta \hbar \omega_S \simeq 10$ for the 
%CD mode.
%On the other hand, the Maradudin-Fein formula, Eq.~(\ref{eq:MFformula}), becomes 
%\begin{eqnarray}
%W &=& W_{\rm decay} + W_{\rm coll},
%\label{eq:MFformula2}
%\\
%W_{\rm decay} 
%&=&
%\frac{\hbar}{2 m_S \omega_S} 
%\sum_{k,k'} \frac{(A_{k,k'}^{(2)})^2}{\omega_k \omega_{k'}}
%\frac{\gamma (1+n_k +n_{k'})}{\gamma^2 +(\omega_S -\omega_k-\omega_{k'})^2},
%\\
%W_{\rm coll}
%&=&
%\frac{\hbar}{m_S \omega_S} 
%\sum_{k,k'} \frac{(A_{k,k'}^{(2)})^2}{\omega_k \omega_{k'}}
%\frac{\gamma (n_k -n_{k'})}{\gamma^2+(\omega_S +\omega_k-\omega_{k'})^2}.
%\end{eqnarray}
We employ this expression
%Eqs.~(\ref{eq:VERrate3}) and (\ref{eq:MFformula2}) 
in our subsequent calculations.

\subsection{Results}

\subsubsection{Classical calculation}

\begin{figure}[htbp]
\hfill
\begin{center}
\begin{minipage}{.32\linewidth}
\includegraphics[scale=0.60]{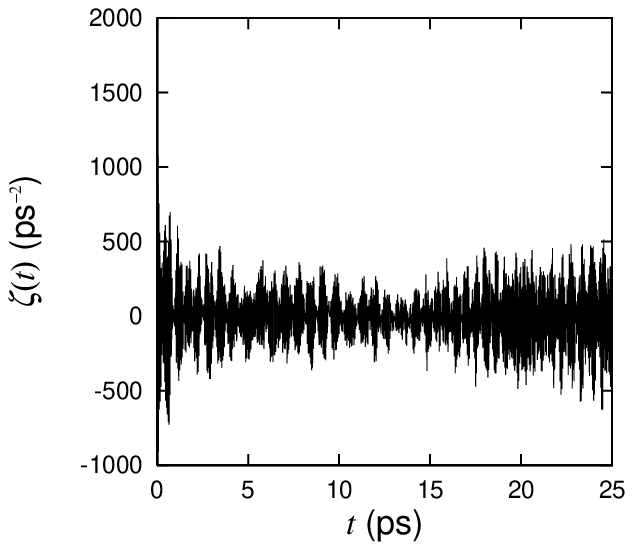}
\end{minipage}
%\hspace{1cm}
\begin{minipage}{.32\linewidth}
\includegraphics[scale=0.60]{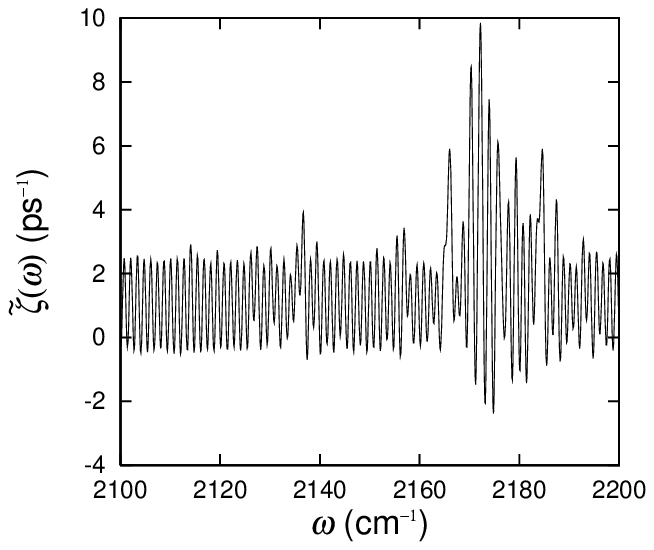}
\end{minipage}
\begin{minipage}{.32\linewidth}
\includegraphics[scale=0.60]{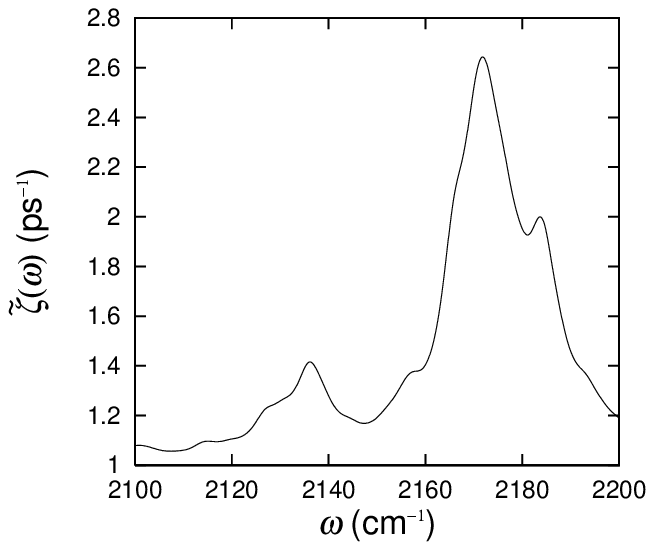}
\end{minipage}
\begin{minipage}{.32\linewidth}
\includegraphics[scale=0.60]{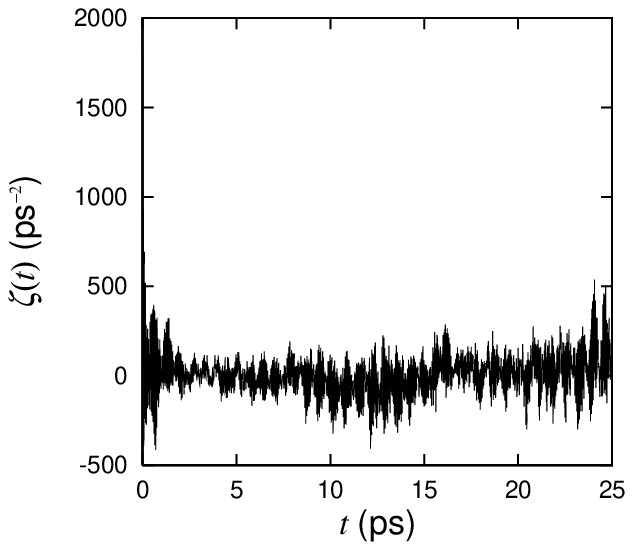}
\end{minipage}
%\hspace{1cm}
\begin{minipage}{.32\linewidth}
\includegraphics[scale=0.60]{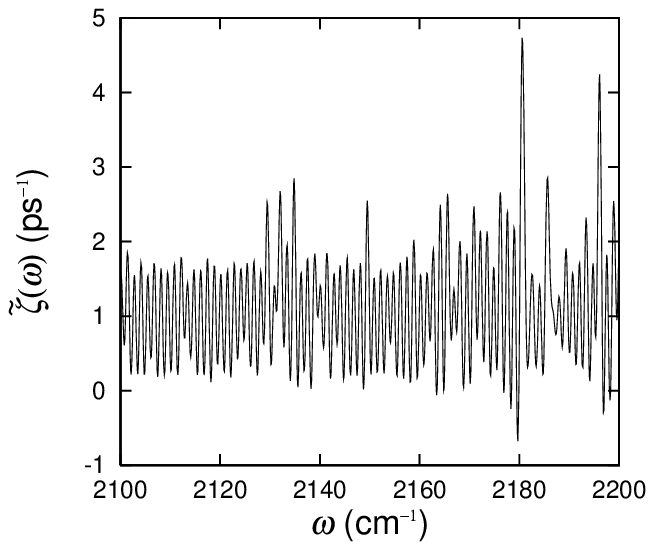}
\end{minipage}
\begin{minipage}{.32\linewidth}
\includegraphics[scale=0.60]{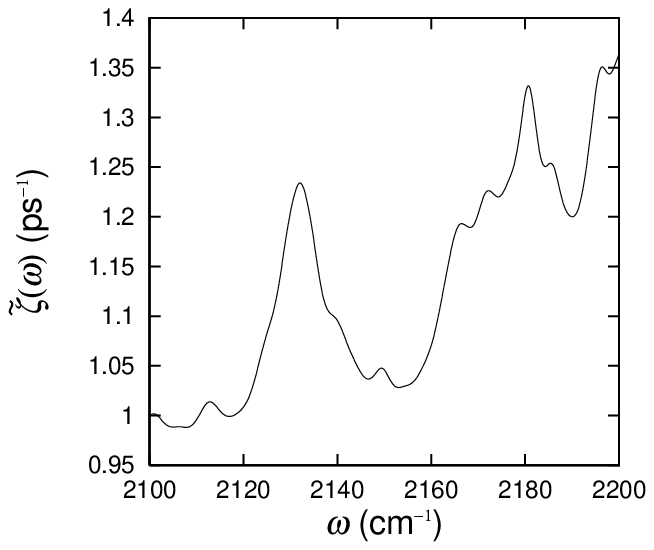}
\end{minipage}
\begin{minipage}{.32\linewidth}
\includegraphics[scale=0.60]{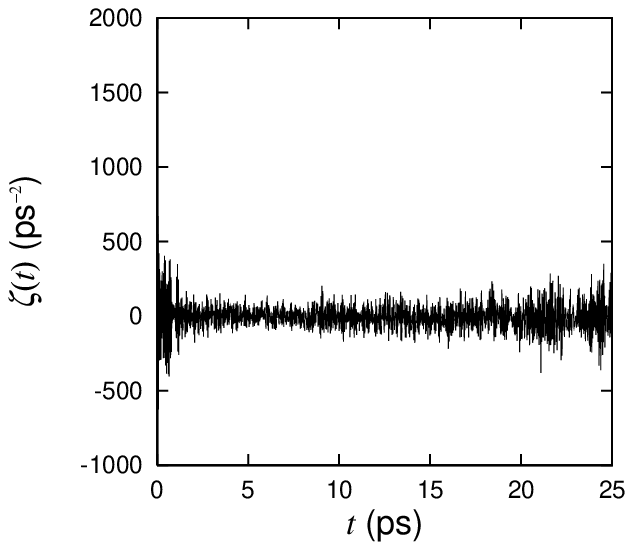}
\end{minipage}
%\hspace{1cm}
\begin{minipage}{.32\linewidth}
\includegraphics[scale=0.60]{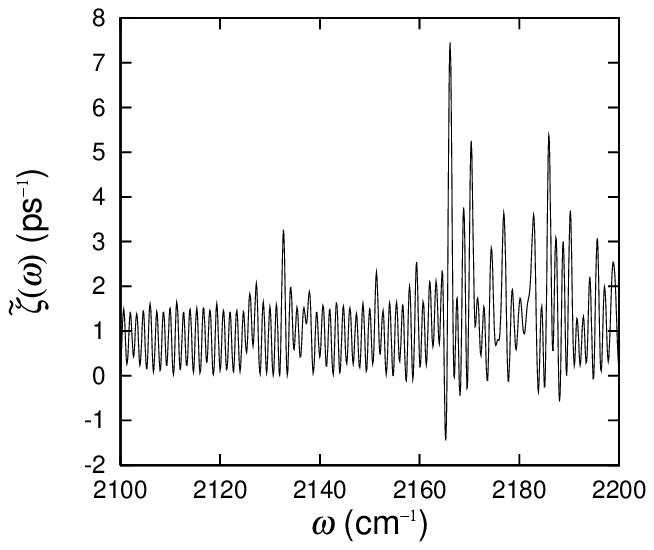}
\end{minipage}
\begin{minipage}{.32\linewidth}
\includegraphics[scale=0.60]{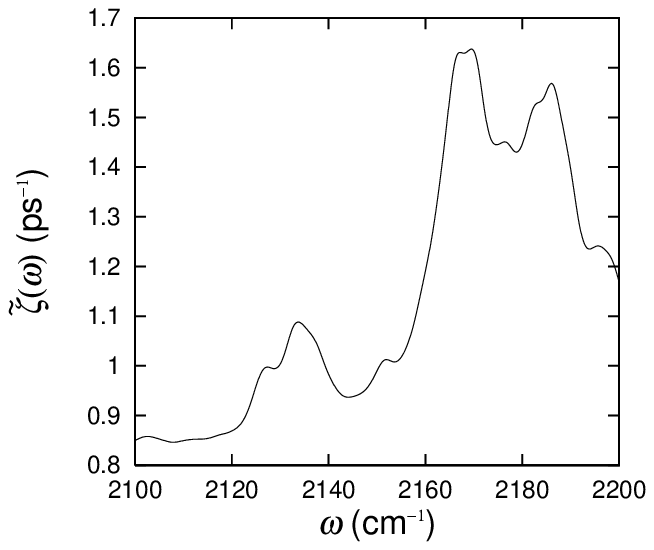}
\end{minipage}
\begin{minipage}{.32\linewidth}
\includegraphics[scale=0.60]{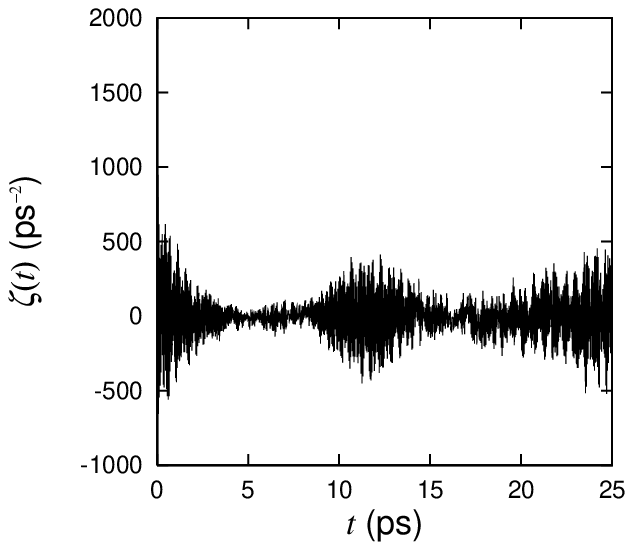}
\end{minipage}
%\hspace{1cm}
\begin{minipage}{.32\linewidth}
\includegraphics[scale=0.60]{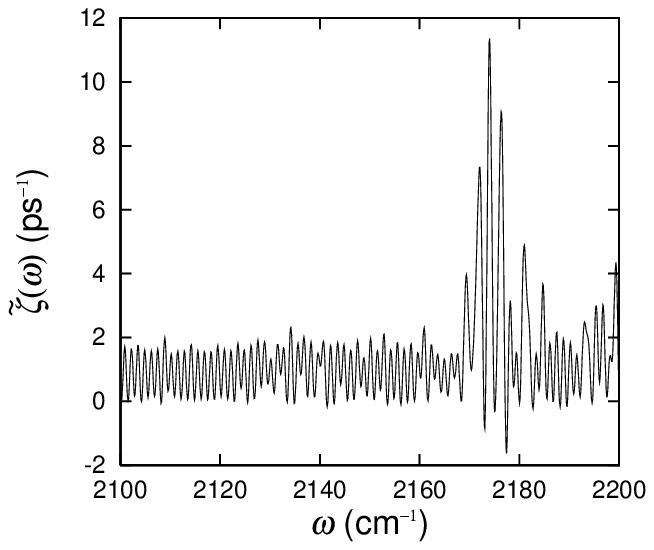}
\end{minipage}
\begin{minipage}{.32\linewidth}
\includegraphics[scale=0.60]{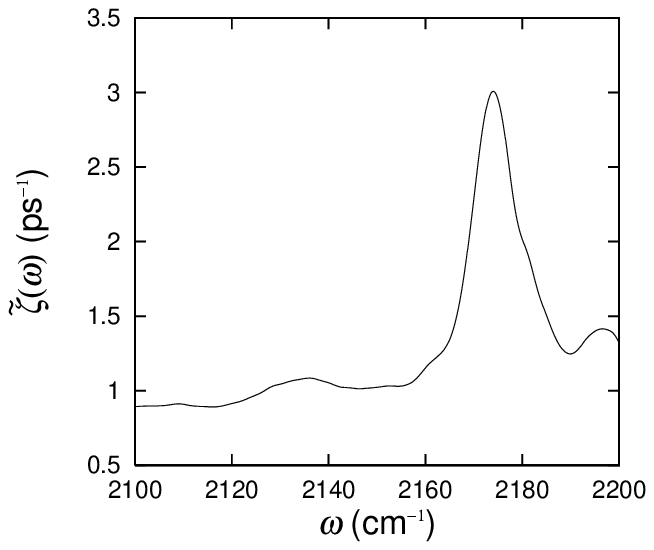}
\end{minipage}
\begin{minipage}{.32\linewidth}
\includegraphics[scale=0.60]{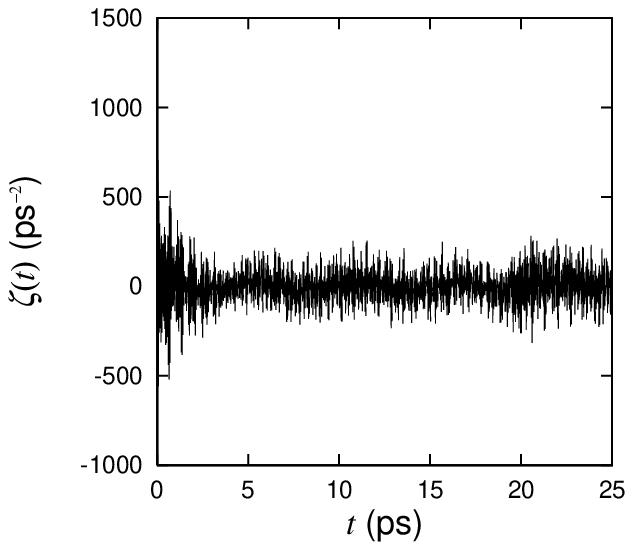}
\end{minipage}
%\hspace{1cm}
\begin{minipage}{.32\linewidth}
\includegraphics[scale=0.60]{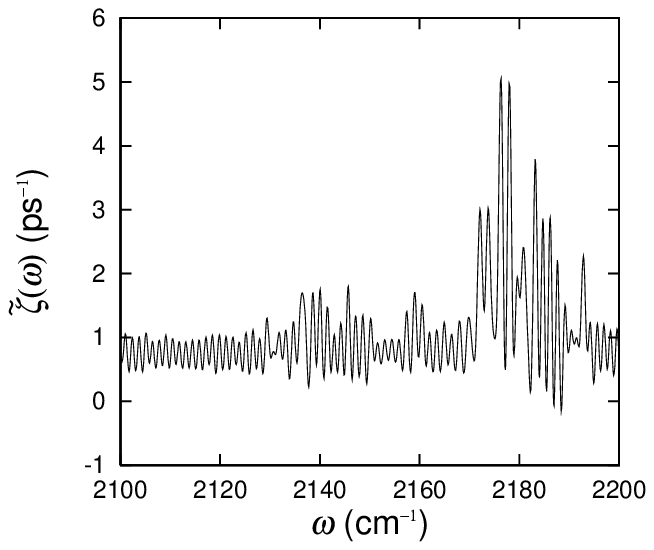}
\end{minipage}
\begin{minipage}{.32\linewidth}
\includegraphics[scale=0.60]{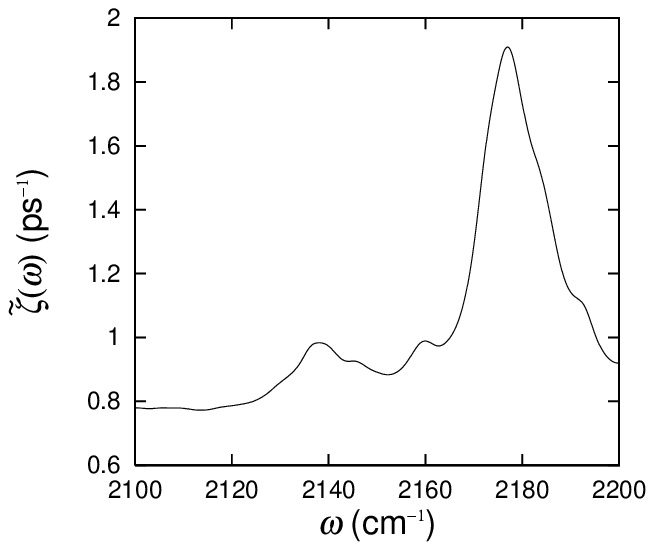}
\end{minipage}
\end{center}
\caption{
%{\sf 
Left: Classical data for the force-force correlation function.
Middle: Fourier spectra for the correlation function. 
Right: The corresponding coarse-grained Fourier spectra.
The ``lifetime'' width parameter $\gamma=3$ cm$^{-1}$.
%}
}
\label{fig:classical}
\end{figure}

Classical force-force correlation functions and 
their (cosine) Fourier transformations are shown 
in the left and middle 
column of Fig.~\ref{fig:classical} for five different trajectories.
Here we have defined a $\zeta$ function as 
\begin{equation}
\zeta(t)=\frac{\beta}{m_S} S_{\rm cl}(t),
\end{equation}
and its (cosine) Fourier transformation as $\tilde{\zeta}(\omega)$, i.e.,
$1/T_1^{\rm cl} = \tilde{\zeta}(\omega_S)$.
Note that these data are obtained from molecular dynamics 
simulations of cyt c in water \cite{BS03}.
As can be seen, the correlation functions oscillate wildly, 
and the (cosine) Fourier transformations are messy.
As such, it is difficult to extract a reliable and stable 
value for the VER rate.

To address this problem,
we introduce the window function 
%In the case of cyt c, $\mu=1.7$ and $k_B T =0.6$ for $T=300$K, 
%so $\mu k_B T \simeq 1.0$ (AKMA), $\beta \hbar \omega_S \simeq 10$,
%and $S(t) \simeq \zeta(t)$. 
%Because $\zeta(t)$ is highly oscillating, we introduce {\bf a width function}:
\begin{equation}
w(t)=\exp(-\gamma t).
\label{eq:window}
\end{equation}
The $\zeta$ functions are multiplied by this function and (cosine) Fourier transformed.
This corresponds to broadening each peak of a spectrum with a Lorentzian with width $\gamma$.
The results for five trajectories are shown in the right column of Fig.~\ref{fig:classical}.
(The width parameter is taken as $\gamma=3$ cm$^{-1}$.)
The results in the right column are better behaved than those in 
the middle column but there still remain some fluctuations.

According to Bu and Straub simulations of cyt c in water \cite{BS03},
we take $\omega_S=2135$ cm$^{-1}$
to investigate the $\gamma$ dependence of the result as shown in 
the left of Fig.~\ref{fig:gamma}.
We see that  
$\tilde{\zeta}(\omega_S) \simeq 1.1 \sim 1.2$ ps$^{-1}$ for $\gamma \simeq 
3 \sim 30$ cm$^{-1}$.
Since $Q(\omega_S)/(\beta \hbar \omega_S) \simeq 2.4 \sim 3.0$ for two-phonon
processes \cite{BS03}, this corresponds to a VER time of $0.3 \sim 0.4$ ps according 
to Eq.~({\ref{eq:qcf}).\footnote{In the VER calculation of myoglobin \cite{Leitner01}, 
Leitner's group took $\gamma = 0.5 \sim 10$ cm$^{-1}$ to be the width, 
and confirmed that the result is relatively insensitive to the 
choice of $\gamma$ in this range.}

\subsubsection{Quantum calculation}

We use the formula Eq.~(\ref{eq:VERrate3})
as a quantum mechanical estimate of the VER rate.
The $\gamma$ dependence of the result is shown on 
the right hand side of Fig.~\ref{fig:gamma}.
We see that, for $\gamma \simeq 3 \sim 30$ cm$^{-1}$, 
the quantum mechanical estimate gives $T_1 \simeq 0.2 \sim 0.3$ ps, 
which is similar to the classical estimate Eq.~(\ref{eq:qcf}):
$T_1 \simeq 0.3 \sim 0.4$ ps.

\begin{figure}[htbp]
\hfill
\begin{center}
\begin{minipage}{.42\linewidth}
\includegraphics[scale=1.0]{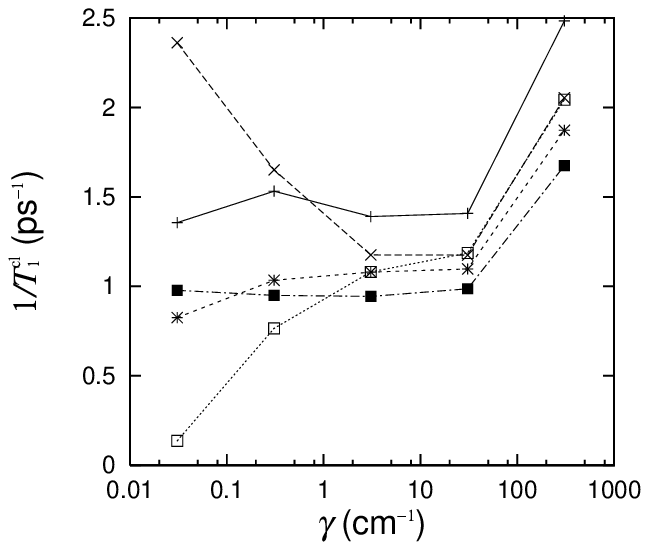}
\end{minipage}
\hspace{1cm}
\begin{minipage}{.42\linewidth}
\includegraphics[scale=1.0]{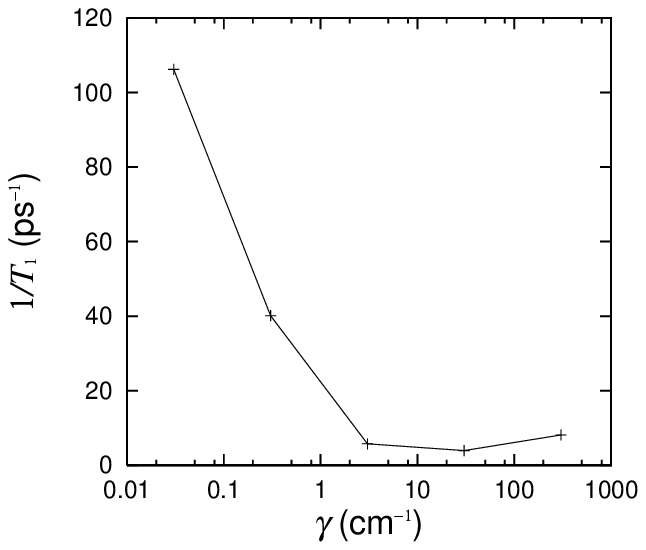}
\end{minipage}
\end{center}
\caption{
%{\sf 
Left: Classical VER rate 
for five trajectories as a function of the ``lifetime'' width 
parameter $\gamma$.
Right: 
VER rate calculated by Eq.~(\ref{eq:VERrate3})
as a function of $\gamma$. 
%}
}
\label{fig:gamma}
\end{figure}

%Good things about these formulas are that we can analyze the 
%result in detail and we can calculate temperature dependence of 
%the result rather easily.
In Tables \ref{tab:contribution1}, %and \ref{tab:contribution2},
we list the largest contributions to the VER rate for 
different width parameters.
For the case of $\gamma=3$ cm$^{-1}$, 
the largest contribution is due to modes (3823,1655).
This combination of modes is nearly resonant with the CD mode
as $|\omega_{3823}+\omega_{1655}-\omega_{CD}| \simeq 0.03$ cm$^{-1}$.
Though the coupling element  for the combination 
is small ($|A^{(2)}_{3823,1655}| =5.1$ kcal/mol/\AA$^3$),
this mode combination contributes significantly to the VER rate.
On the other hand, 
for the case of $\gamma=30$ cm$^{-1}$, 
the largest contribution results from 
the combination of modes (3330,1996).
This combination is somewhat off-resonant, 
i.e., 
$|\omega_{3330}+\omega_{1996}-\omega_{CD}|=32$ cm$^{-1}$,
but the coupling element is very large 
($|A^{(2)}_{3330,1996}| =22.3$ kcal/mol/\AA$^3$),
and the contribution is significant.
In both cases, one combination of modes dominates the VER rate ($\simeq 20$\%) 
though there are non-negligible contributions from other combinations of modes.

\begin{table}[htbp]
\caption{
%{\sf 
The largest contributions to the VER rate (in units of ps$^{-1}$) for
%The width parameter is $\gamma=3$ cm$^{-1}$.
$\gamma=3$ cm$^{-1}$ (left) and 
$\gamma=30$ cm$^{-1}$ (right).
Note that $(k,l)$ are mode numbers, not wavenumbers.
%}
}
\hfill
\begin{center}
\begin{tabular}{c|c}
\hline \hline
$(k,l)$ & contribution \\ \hline \hline
(3823, 1655) & 1.10 ( 19 \%) \\ \hline
(3823, 1654) & 0.43 ( 8 \%) \\ \hline
(3822, 1655) & 0.37 ( 6 \%) \\ \hline
(3330, 1996) & 0.17 ( 3 \%) \\ \hline
(3822, 1654) & 0.15 ( 3 \%) \\ \hline
(3823, 1661) & 0.14 ( 3 \%) \\ \hline
(3822, 1661) & 0.05 ( 1 \%)\\ \hline 
(3822, 1656) & 0.05 ( 1 \%)\\ \hline
(3823, 1658) & 0.04 ( 1 \%)\\ 
\hline
\end{tabular}
\hspace{2cm}
\begin{tabular}{c|c}
\hline \hline
$(k,l)$ & contribution \\ \hline \hline
(3330, 1996) & 0.88 (22 \%)\\ \hline
(3823, 1655) & 0.11 (3 \%)\\ \hline
(3170, 2196) & 0.07 (2 \%)\\ \hline
(1996, 1996) & 0.05 (1 \%)\\ \hline
(3823, 1654) & 0.04 (1 \%)\\ \hline
(3170, 2202) & 0.04 (1 \%)\\ \hline
(3822, 1655) & 0.04 (1 \%)\\ \hline
(3327, 1996) & 0.03 (1 \%)\\ \hline
(3330, 1655) & 0.02 (1 \%)\\ %\hline
%(3823, 1661) & 0.02 (1 \%)\\ 
\hline
\end{tabular}
\end{center}
\label{tab:contribution1}
\end{table}

We have also examined the temperature dependence of the VER rates using 
Eq.~(\ref{eq:VERrate3}).
As shown in Fig.~\ref{fig:temp},
for $T <300$ K there is little temperature dependence as has been 
addressed in the case of myoglobin \cite{Leitner01}.
Thus we can say that the relaxation of the CD mode is quantum mechanical 
rather than thermal because the decay at 300 K is similar to 
that at 0.3 K.
%In this case, both formulas, 
%Eqs.~(\ref{eq:VERrate3}) and (\ref{eq:MFformula2}), 
%give the similar results again.

\begin{figure}[htbp]
\hfill
\begin{center}
%\begin{minipage}{.42\linewidth}
\includegraphics[scale=1.0]{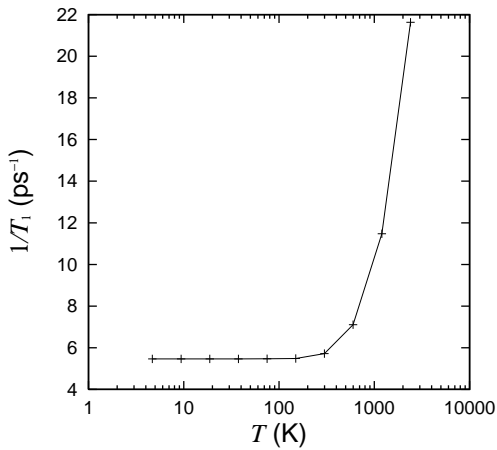}
\end{center}
\caption{
%{\sf 
Temperature dependence of the VER rate 
calculated by Eq.~(\ref{eq:VERrate3}). 
The width parameter is $\gamma=3$ cm$^{-1}$.
%}
}
\label{fig:temp}
\end{figure}

%\begin{table}[htbp]
%\caption{
%{\sf 
%Temperature dependence of the VER times. 
%The width parameter is $\gamma=3$ cm$^{-1}$.
%}}
%\hfill
%\begin{center}
%\begin{tabular}{c|c|c}
%\hline \hline
%$T$ (K) & perturbation (ps) & MF formula (ps) \\ \hline \hline
%0.3 & 0.18 & 0.18 \\ \hline
%3   & 0.18 & 0.18 \\ \hline
%30  & 0.18 & 0.18 \\ \hline
%300 & 0.17 & 0.18 \\ 
%\hline
%\end{tabular}
%\end{center}
%\label{tab:temp}
%\end{table}

\subsubsection{Discussion}

We examine the relationship between the theoretical results 
described above and the corresponding experiments of Romesberg's group 
which has studied the spectroscopic 
properties of the CD mode in cyt c \cite{CJR01}. 
They measured the shifts and widths of the spectra 
for different forms of cyt c;
the widths of the spectra (FWHM) 
were found to be $\Delta \omega_{\rm FWHM} \simeq 6.0 \sim 13.0$ cm$^{-1}$.
A rough estimate of the VER rate leads to 
\begin{equation}
T_1 \sim 5.3/\Delta \omega_{\rm FWHM}  \,({\rm ps})
\label{eq:estimate}
\end{equation}
which corresponds to $T_1 \simeq 0.4 \sim 0.9$ ps.
This estimate is similar to the ``semi''-classical prediction 
computed using Eq.~(\ref{eq:qcf}) and appropriate QCFs (0.3 $\sim$ 0.4 ps)
and the perturbative quantum mechanical estimate using the reduced model (0.2 $\sim$ 0.3 ps).
This result appears to justify the use of QCFs and the reduced model 
in this situation, and suggests that the effects of the protein solvation (by water) 
are negligible in describing the VER of the CD mode. 
Of course, we must be careful in comparing 
the estimate derived from (\ref{eq:estimate}) as 
there may be inhomogeneity in the experimental 
spectra.\footnote{We have confirmed that the methyl group does not 
rotate during the equilibrium simulations. Thus we can exclude 
the rotation as a possible reason of inhomogeneity.} 
As such, it is more 
desirable to calculate not only VER rates but 
spectroscopic observables themselves to compare with experiments.

Finally we discuss the relation between this work and previous 
work on carbon-monoxide myoglobin (MbCO).
Though there are many experimental studies on Mb \cite{MK97},
we focus on the experiments of Anfinrud's group \cite{Anfinrud} and 
Fayer's group \cite{Fayer96} on MbCO.
The former group found that the VER time for CO in the heme pocket (photolyzed MbCO) 
is $\simeq 600$ ps, whereas the latter group found 
that the VER time for CO bounded to the heme is $\simeq 20$ ps.
This difference is interpreted as follows: 
CO is covalently bonded to the heme for the latter case, 
whereas CO is ``floating'' in the pocket in the former case, 
i.e., the force applied to CO for the latter case is stronger 
than that for the former case. This difference in the magnitude of the force causes 
the slower VER for the ``floating'' CO.
In this respect, the CD bond is expected to be stronger 
than the CO-heme coupling.
This may explain a VER time $\sim 0.1$ ps, 
which is similar to the VER times for the CH(CD) stretching 
modes in benzene (or perdeuterobenzene) \cite{SHR84,Schranz96}. 
It will be interesting to apply a similar reduced model
to the analysis of the VER of CO in MbCO. 
%The analysis of heme cooling \cite{BS03b} using this 
%type of reduced model also should be pursued. 

\section{Summary and further aspects}
\label{sec:summary}

After reviewing the VER rate formula derived from quantum mechanical 
perturbation calculations, 
we applied it to the analysis of VER of a CD stretching mode in cyt c.
We modeled cyt c in vacuum 
as a normal mode system with the third-order anharmonic coupling elements,
which were calculated from the CHARMM potential. 
We found that, for the width parameter $\gamma=3 \sim 30$ cm$^{-1}$, 
the VER time is $0.2 \sim 0.3$ ps, which agrees rather well 
with the previous classical calculation using the quantum correction factor (QCF)
method, and is consistent with the experiments by Romesberg's group.
This result indicates that the use of QCFs or a reduced model Hamiltonian 
can be justified a posteriori to describe the VER problem.
We decomposed the VER rate into contributions from two modes, 
and found that the most significant 
contribution, which depends on the ``lifetime'' width parameter, 
results from modes most resonant with the CD mode.
%We also found that the VER rate derived here, which is equivalent to 
%that derived by Shiga and Okazaki, behaves very similarly with the 
%Maradudin-Fein formula in the case of the CD mode in cyt c.
%\\

Finally we note several future directions which should be studied:
(a) Our final results for the VER rate depend on a width parameter $\gamma$.
Unfortunately we do not know which value is the most appropriate for $\gamma$. 
Non-equilibrium simulations (with some quantum corrections \cite{NS03}) might help this 
situation, and are useful to investigate energy path ways or sequential IVR (intramolecular 
vibrational energy redistribution) \cite{SF98} in a protein.
(b) This work is motivated by pioneering spectroscopic experiments by Romesberg's group.
The calculation of the VER rate and the  linear or nonlinear response 
functions, related to absorption or 2D-IR (or 2D-Raman) 
spectra \cite{OT97,JSD00,MNTALF03,EH03}, is desirable.
(c) Romesberg's group investigated a spectroscopic change due to  
the oxidation or reduction of Fe in the heme; 
such an electron transfer process \cite{ET} is fundamental for the functionality of cyt c.
To survey this process dynamically, it will be necessary to combine some quantum chemistry 
(ab initio) calculations with MD simulations \cite{DFT,QMMM,Komeiji}.

%\begin{acknowledgments}
The authors thank 
Dr.~J.~Gong for noting Ref.~\cite{KTF94},
Prof.~A.~Kidera, Prof.~S. Okazaki,
Prof.~J.~L.~Skinner,
Prof.~D.M.~Leitner, Prof.~Y. Mizutani, 
Dr.~T. Takami, Dr.~T. Miyadera,
Dr.~Y. Kawashima, 
Dr.~S.~Fuchigami,
Mr.~H. Teramoto for useful discussions,
Prof.~A. Stuchebrukhov for sending reprints,
and Dr.~M. Shigemori for providing perl scripts used in this work.
%\end{acknowledgments}

\end{document}